\def\FLATarxiv{1}
\documentclass[conference]{IEEEtran}
\IEEEoverridecommandlockouts

\usepackage{cite}
\usepackage[T1]{fontenc}
\usepackage[final]{microtype}
\usepackage{indentfirst}
\usepackage{multirow}
\usepackage{amsmath,amssymb,amsfonts}
\usepackage{graphicx}
\graphicspath{{./figure/}}
\usepackage{array}
\usepackage{tabularx}
\usepackage{booktabs}
\usepackage{makecell}
\usepackage[table]{xcolor}
\newcommand{\std}[1]{{\tiny\color{gray!90}$\pm$#1}}
\newcommand{\tjb}[1]{\textbf{#1}}
\newcommand{\flatfull}{\textbf{\underline{F}}ingerprint-guided \textbf{\underline{L}}earnable \textbf{\underline{A}}cquisition for \textbf{\underline{T}}raffic Calibration}
\newcolumntype{L}{>{\raggedright\arraybackslash}X}
\newcolumntype{C}{>{\centering\arraybackslash}X}
\usepackage{url}
\usepackage{xurl}
\usepackage{float}
\usepackage[caption=false,font=footnotesize,labelfont=rm]{subfig}
\usepackage{tikz}
\usepackage{pgfplots}
\pgfplotsset{compat=1.18}
\usepackage{algorithm}
\usepackage{algpseudocode}

\def\BibTeX{{\rm B\kern-.05em{\sc i\kern-.025em b}\kern-.08em
    T\kern-.1667em\lower.7ex\hbox{E}\kern-.125emX}}
\ifdefined\pdfminorversion
\fi

\newcommand{\subcap}[1]{{\fontfamily{ptm}\selectfont #1}}

\newcommand{\plotpanelpair}[1]{\includegraphics[width=0.495\linewidth,keepaspectratio]{#1}}
\newcommand{\plotpanelcomposite}[1]{\includegraphics[width=0.49\linewidth,keepaspectratio]{#1}}
\newcommand{\panelh}{\dimexpr0.46\columnwidth*9/16\relax}
\newcommand{\sceneimgpair}[1]{\includegraphics[width=0.47\linewidth,height=\panelh]{#1}}
\newcommand{\panelwfull}{\columnwidth}

\newcommand{\plotpanelfull}[1]{\includegraphics[width=\panelwfull,keepaspectratio]{#1}}

\title{FLAT: Smoothing the Rugged Landscape for Learnable, Sample-Efficient Traffic Calibration}

\ifdefined\FLATarxiv
\author{\IEEEauthorblockN{Haopeng Deng\thanks{Submitted to the 2027 IEEE Intelligent Vehicles Symposium (IV). Code and cache: \protect\url{https://github.com/QianyHP/FLAT-calibration}.}}
\IEEEauthorblockA{\textit{School of Future Transportation} \\
\textit{Guangzhou Maritime University}\\
Guangzhou 510725, China \\
qianyhp@gmail.com}
\and
\IEEEauthorblockN{Thomas He}
\IEEEauthorblockA{\textit{BGP Inc.} \\
Hebei 072751, China \\
thomasho3860589@outlook.com}
\and
\IEEEauthorblockN{Dayuan Wang}
\IEEEauthorblockA{\textit{Department of Biostatistics} \\
\textit{University of Florida}\\
Gainesville, FL 32601, USA \\
dayuan.wang@ufl.edu}}
\else
\author{}
\fi

\begin{document}
\bstctlcite{IEEEAuthorControl}
\maketitle
\begin{abstract}
Calibrating microscopic traffic models for urban digital twins is an expensive black-box optimization problem: evaluating each set of car-following and lane-changing parameters requires a full simulation, limiting the evaluations available per recalibration window. Matching raw trajectories yields a rugged objective that sparse surrogates struggle to learn, reducing sequential acquisition to near-random probing. We present FLAT, which couples \emph{what} to optimize with \emph{where} to sample next. An eight-dimensional behavioral fingerprint smooths the parameter--error landscape, making the objective learnable from a few dozen samples; annealed lower-confidence-bound (LCB) acquisition then spends each remaining run where it most reduces error. The LCB loop supports a Gaussian process (GP), random forest (RF), or multi-layer-perceptron (MLP) ensemble. Across six heterogeneous real-world scenes, FLAT-GP achieves the lowest scene-averaged behavioral error, winning 6/6 scenes against SPSA, GA, and CMA-ES and 5/6 against TPE under the matched 100-run budget. Some baselines require up to $4.4$ times as many simulations to match. Ablations show objective choice shifts final behavioral error by $80.6\%$ on average, removing sequential LCB raises the six-scene mean by $21.7\%$, and surrogate choice shifts it by at most $4.2\%$, confirming gains trace to objective geometry and sequential allocation rather than surrogate capacity.
\end{abstract}


\section{Introduction}
\label{sec:introduction}

Microscopic traffic digital twins support simulation-based evaluation and traffic control~\cite{ref_dt}. Their credibility rests on a microscopic simulator such as SUMO (Simulation of Urban MObility)~\cite{ref_sumo} reproducing how vehicles move, not merely average flow; fidelity then depends on car-following~\cite{ref_krauss} and lane-changing~\cite{ref_lc2013} parameters calibrated per location from trajectories. As observed traffic conditions change, those parameters require periodic refresh. Each parameter setting requires a full network simulation, making calibration an expensive, stochastic, black-box optimization problem under a limited budget per recalibration window~\cite{ref_huainan2022}.

Two schools address this problem. The first searches the raw objective directly with general-purpose optimizers, including genetic algorithms (GA)~\cite{ref_ga}, simultaneous perturbation stochastic approximation (SPSA)~\cite{ref_spsa}, and the covariance matrix adaptation evolution strategy (CMA-ES)~\cite{ref_cmaes}. They are derivative-free but sample-hungry; under a hundred-run budget, an evolution strategy completes only a handful of generations. The second is surrogate-assisted, or Bayesian, optimization, which fits a cheap response-surface model to decide where to evaluate next. Jones~\textit{et~al.}~\cite{ref_surrogate2} formalized the idea, and the approach is now standard for costly simulators~\cite{ref_surrogate1}, including the tree-structured Parzen estimator (TPE)~\cite{ref_tpe}, Gaussian-process (GP)~\cite{ref_pracbo}, random-forest (RF)~\cite{ref_smac2011}, and deep-learning multi-layer-perceptron (MLP) ensemble surrogates. The same paradigm now extends to large-scale traffic-model calibration~\cite{ref_bocal,ref_aical,ref_bo4mob}. Yet no single optimizer wins consistently across setups~\cite{ref_calibration}, leaving open how much efficiency depends on the objective rather than the optimizer alone.

Surrogate guidance depends on an objective learnable from few samples, yet in microscopic traffic calibration, trajectory-derived errors can be rugged and scene-dependent, hindering reliable surrogate learning. Pointwise errors can amplify small timing shifts, while network averages can conceal differences in speed dispersion and stop--go behavior. Two coupled choices shape sample efficiency: the \emph{geometry of the objective} a surrogate must learn, and \emph{how a fixed budget is split} between initial exploration and sequential acquisition. Without a learnable objective, acquisition has no reliable basis for ranking unevaluated parameter settings.

\begin{figure*}[t]
  \centering
  \includegraphics[width=\textwidth,keepaspectratio]{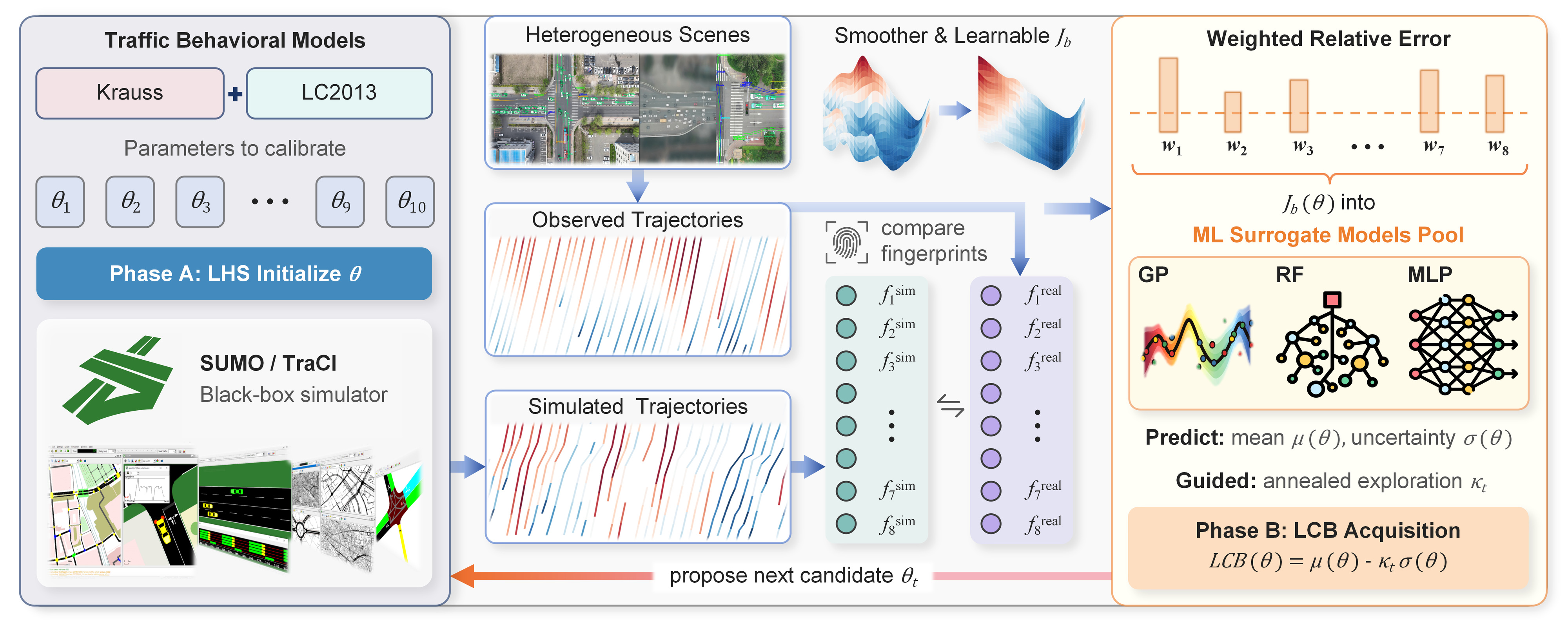}
  \caption{FLAT overview. Black-box calibration of Krauss car-following and LC2013 lane-changing parameters in SUMO--TraCI on heterogeneous scenes. Behavioral fingerprints encode trajectories into a smoother parameter--error landscape, making $J_b(\theta)$ learnable for sample-efficient acquisition. Phase~A explores $\theta$ by Latin-hypercube sampling (LHS); Phase~B refits a GP, RF, or MLP surrogate and selects the next evaluation by annealed LCB.}
  \label{fig:arch}
\end{figure*}

We instantiate this as \textbf{FLAT} (\flatfull), which makes both choices explicit. For the objective, it replaces trajectory matching with a compact \emph{behavioral fingerprint} that smooths the parameter--error landscape, enabling reliable surrogate learning from a few dozen runs. For the budget, an uncertainty-aware acquisition policy allocates the remaining simulations sequentially, annealing exploration into exploitation as the budget depletes. The framework is surrogate-agnostic, which lets us test whether efficiency comes from model capacity or from how the problem is posed. Fig.~\ref{fig:arch} overviews the framework.

\vspace{3pt}
\textbf{Contributions.}
\begin{enumerate}
  \item We propose FLAT, a surrogate-acquisition framework that converts the rugged objective into a learnable surface via behavioral representation reshaping, enabling sample-efficient calibration within $10^2$ evaluations per scene.
  \item We establish a consistent empirical advantage over four mechanistically diverse optimizer families across six real-world scenes, with some competitors requiring up to $4.4$ times more simulations to reach equivalent quality.
  \item We show through ablations that objective geometry and sequential allocation, not surrogate capacity, drive the gain; objective choice shifts error by $80.6\%$, sequential allocation by $21.7\%$, surrogate choice by at most $4.2\%$.
\end{enumerate}

\section{Literature Review}
\label{sec:related}

Calibration objectives determine both what behavior is matched and how mismatch is penalized. Punzo~\textit{et~al.}~\cite{ref_punzo2021} use Pareto-efficiency and indifference curves to expose trade-offs in fitting spacing, speed, and acceleration, while Schrader~\textit{et~al.}~\cite{ref_schrader2024} calibrate car-following models against radar trajectories in SUMO. Moving to distributional mismatch, Abuamer and Tamp\`ere~\cite{ref_abuamer2024} compare Wasserstein distance, Kolmogorov--Smirnov distance, and root mean squared relative error for speed and headway calibration under SPSA and Bayesian optimization. Their results favor objectives that combine behavioral indicators and preserve distributional information, improving calibration efficiency and parameter convergence consistency. These choices also define the response surface a calibration surrogate must learn.

Sample-efficient calibration can exploit traffic structure as well as statistical response models. Zhang~\textit{et~al.}~\cite{ref_zhang2017} combine simulated observations with an analytical network model in a metamodel, using structural information to guide large-scale calibration. An analytical component supplies network structure, while a fitted correction adapts it to simulated observations. Sha~\textit{et~al.}~\cite{ref_sha2020} compare Bayesian acquisition with SPSA for SUMO calibration. They report acceptable speed-distribution fit for both methods at multiple sensor locations, with BO converging more efficiently. More recently, BO4Mob~\cite{ref_bo4mob} benchmarks scalable optimization for high-dimensional origin--destination demand estimation from traffic sensors. These approaches optimize a specified loss; FLAT also examines how objective design affects surrogate learning.

Representation learning provides a further connection between problem formulation and efficient search. Moriconi~\textit{et~al.}~\cite{ref_moriconi2020} jointly learn a low-dimensional parameter embedding, a response surface, and a reconstruction map, allowing Bayesian acquisition to operate in a learned subspace. LANCER~\cite{ref_lancer2023} instead learns a smooth surrogate of a solver--objective composition, providing gradients for coupled learning and optimization without differentiating through the solver. FLAT places the representation at the simulator output: fixed behavioral summaries define the mismatch before search, without fitting an additional encoder. We test whether this objective can be learned from few evaluations and guide sequential acquisition, using landscape diagnostics, held-out prediction, and equal-budget experiments.

\section{Problem Formulation}
\label{sec:problem}

Calibration tunes six Krauss car-following~\cite{ref_krauss} and four LC2013 lane-changing~\cite{ref_lc2013} parameters in SUMO~\cite{ref_sumo} via TraCI, the traffic control interface, as a ten-dimensional vector $\theta$ set at runtime; physical meanings in Appendix~\ref{app:params}. Trajectory matching scores spatiotemporal sequences \emph{pointwise}:
\begin{equation}
  J_{\mathrm{traj}}(\theta)=\frac{1}{|\mathcal{V}|}\sum_{v\in\mathcal{V}}\frac{1}{T}\sum_{t=1}^{T}\bigl\|\mathbf{y}_v^{\mathrm{sim}}(\theta,t)-\mathbf{y}_v^{\mathrm{real}}(t)\bigr\|_2^2,
  \label{eq:traj}
\end{equation}
where $\mathbf{y}_v(t)$ denotes the simulated or observed state (e.g.\ position and speed) of vehicle $v\in\mathcal{V}$ at $t\in\{1,\ldots,T\}$, and one seeks $\theta^\star$ that minimizes $J_{\mathrm{traj}}$.

Yet microscopic traffic is a stochastic, strongly coupled dynamical system, making pointwise trajectory matching fragile. Re-seeding changes vehicle timing and interactions, while pointwise errors can obscure the stop--go structure, speed spreading, and queue dynamics that practical calibration requires. Relaxing to network-mean speed over time,
\begin{equation}
  J_s(\theta)=\sqrt{\frac{1}{T}\sum_{t=1}^{T}\bigl(\bar{v}^{\mathrm{sim}}(\theta,t)-\bar{v}^{\mathrm{real}}(t)\bigr)^2},
  \label{eq:js}
\end{equation}
still leaves a rugged, behaviorally misaligned landscape.

Instead, we compare ensemble distributional statistics. Each trajectory set $\mathcal{T}$ is mapped by a fixed $\phi:\mathcal{T}\!\to\!\mathbb{R}^K$, and we sum the per-component relative errors,
\begin{equation}
  J_{\mathrm{dist}}(\theta)=\sum_{k=1}^{K}\frac{\bigl|f_k^{\mathrm{sim}}(\theta)-f_k^{\mathrm{real}}\bigr|}{\max\bigl(|f_k^{\mathrm{real}}|,\epsilon\bigr)},
  \label{eq:dist}
\end{equation}
where $f_k^{\mathrm{sim}}$ and $f_k^{\mathrm{real}}$ are the $k$th components of $\phi$ for simulated and observed traffic. Because each simulation is expensive, and an online digital twin affords only a limited recalibration window, sample efficiency is the binding constraint.

\section{Methodology}
\label{sec:method}

FLAT's principle: \emph{spend each simulation where it matters most}. Two coupled decisions realize it---what to optimize and where to sample next---via a behavioral-fingerprint objective (\S\ref{sec:objective}) and a sequential acquisition loop with a pluggable surrogate (\S\ref{sec:acq}--\S\ref{sec:surrogate}). Algorithm~\ref{alg:flat} details the steps.

\begin{algorithm}[h]
  \caption{FLAT calibration under simulation budget $B$.}
  \label{alg:flat}
  {\small
  \begin{algorithmic}[1]
    \Require Scene config; real fingerprint $\mathbf{f}^{\mathrm{real}}$; bounds; $N_{\mathrm{init}}$; schedule $\kappa_t$
    \Ensure Calibrated $\theta^\star$ and history $\{(\theta_i, J_b(\theta_i))\}_{i=1}^{B}$
    \State Draw $\{\theta_i\}_{i=1}^{N_{\mathrm{init}}}$ by LHS; evaluate $J_b$ in SUMO \Comment{Phase A}
    \For{$t = N_{\mathrm{init}}+1,\ldots,B$} \Comment{Phase B}
      \State Fit surrogate $(\mu,\sigma)$ on $\{(\theta_i,J_b(\theta_i))\}_{i<t}$
      \State $\theta_{\mathrm{best}} \leftarrow \theta_{\arg\min_{i<t} J_b(\theta_i)}$; sample trust region around $\theta_{\mathrm{best}}$ + global candidates; drop near-duplicates
      \State $\theta_t \leftarrow \arg\min_\theta\, \mu(\theta)-\kappa_t\,\sigma(\theta)$
      \State Evaluate $J_b(\theta_t)$ in SUMO; append
    \EndFor
    \State \Return $\theta^\star=\theta_{\arg\min_{1\leq i\leq B} J_b(\theta_i)}$
  \end{algorithmic}
  }
\end{algorithm}

\subsection{Behavioral Fingerprint Objective}
\label{sec:objective}
Behavioral fingerprinting instantiates \eqref{eq:dist} at $K{=}8$ with $\phi$ drawn from core facility performance measures in the FHWA \textit{Traffic Analysis Toolbox}~\cite{ref_fhwa_moe} and the \textit{Highway Capacity Manual}~\cite{ref_hcm}.
Domain-specified and fixed before search rather than learned from data, $\phi$ spends no evaluation budget on feature fitting.
The fingerprint vector is
\begin{equation}
  \mathbf{f}=\bigl(v_{\mathrm{mean}},\,v_{15},\,v_{85},\,\bar{a}^{+},\,\bar{a}^{-},\,a_{5},\,a_{95},\,s_{\mathrm{stop}}\bigr),
  \label{eq:fvec}
\end{equation}
comprising mean speed and its $15$th/$85$th percentiles, mean acceleration and deceleration magnitudes, $5$th/$95$th acceleration percentiles, and stop fraction.

\begin{figure}[h]
  \centering
  \includegraphics[width=\columnwidth]{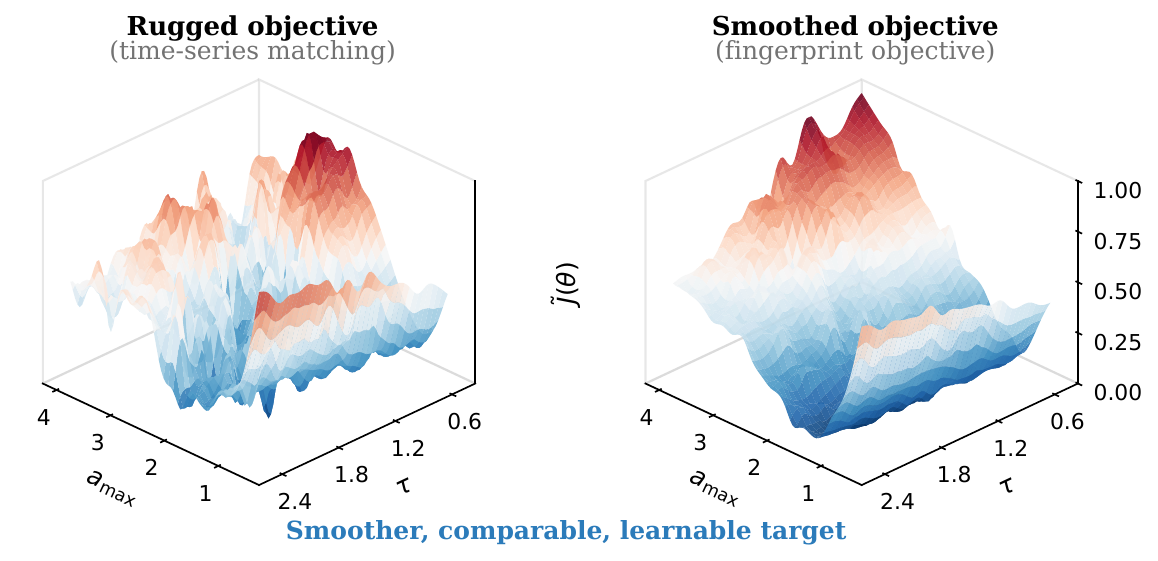}
  \caption{Objective landscapes over maximum acceleration $a_{\max}$ and desired time headway $\tau$ on the XAM expressway, from 400 SUMO runs with others fixed at $\theta^\star$. Surfaces normalized to $[0,1]$. \emph{Left:} $J_s$ \eqref{eq:js}. \emph{Right:} $J_b$ \eqref{eq:jb}.}
  \label{fig:landscape}
\end{figure}

Weights follow inverse-variance weighting (IVW), $w_k \propto 1/\mathrm{Var}_k$ on per-vehicle references, normalized and held fixed; per-term clipping at $3$ refines \eqref{eq:dist} into the objective:
\begin{equation}
  J_b(\theta) = \sum_{k=1}^{8} w_k \,\min\!\left(\frac{|f_k^{\mathrm{sim}}(\theta)-f_k^{\mathrm{real}}|}{\max(|f_k^{\mathrm{real}}|,\epsilon)},\; 3\right).
  \label{eq:jb}
\end{equation}
Fig.~\ref{fig:landscape} contrasts $J_s$ and $J_b$ on the XAM expressway; $J_b$ smooths the parameter--error surface for surrogate guidance.

\subsection{Sequential Acquisition}
\label{sec:acq}
The budget $B$ is split between Phase~A exploration and Phase~B exploitation. Phase~A draws $N_{\mathrm{init}}$ LHS~\cite{ref_lhs} samples over $\theta$ in SUMO. Phase~B spends the remaining $B-N_{\mathrm{init}}$ simulations sequentially: at each step it (i) fits a surrogate returning $\mu(\theta)$ and $\sigma(\theta)$ on observed $(\theta,J_b)$ pairs, (ii) draws candidates from a trust region~\cite{ref_turbo2019} around the incumbent plus a global pool, (iii) selects the candidate minimizing LCB
\begin{equation}
  \mathrm{LCB}(\theta) = \mu(\theta) - \kappa_t\,\sigma(\theta),
  \label{eq:lcb}
\end{equation}
where $\kappa_t$ is the annealed exploration--exploitation weight in \eqref{eq:kappa},
after discarding points too close to evaluated ones, and (iv) evaluates it in SUMO and appends the result.

\begin{figure}[h]
  \centering
  \includegraphics[width=\columnwidth]{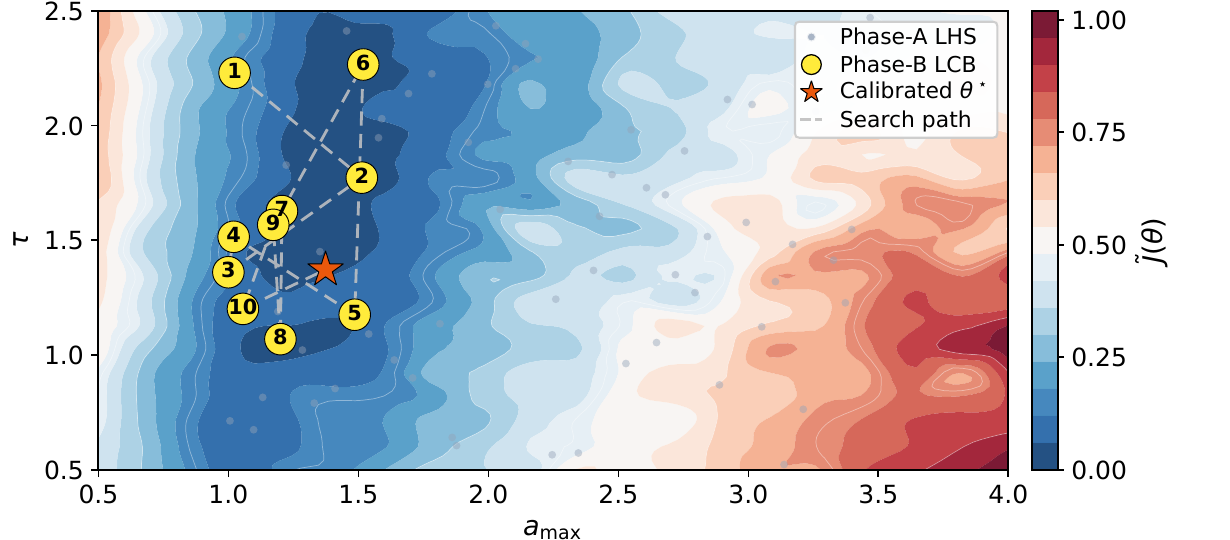}
  \caption{Sequential acquisition projected onto the XAM slice in Fig.~\ref{fig:landscape}; background $J_b$ normalized to $[0,1]$. \emph{Grey:} Phase-A LHS. \emph{Numbered:} selected Phase-B LCB trials in chronological order. \emph{Star:} calibrated $\theta^\star$.}
  \label{fig:lcb}
\end{figure}

The uncertainty term $\sigma$ rewards exploration, while $\mu$ guides exploitation toward low predicted error. Over the $B-N_{\mathrm{init}}$ sequential steps $t=N_{\mathrm{init}}+1,\ldots,B$ we anneal $\kappa_t$ linearly,
\begin{equation}
  \kappa_t = \kappa_0 + (\kappa_T-\kappa_0)\,\frac{t-N_{\mathrm{init}}-1}{B-N_{\mathrm{init}}-1}, \quad \kappa_0{=}2.0,\ \kappa_T{=}0.5,
  \label{eq:kappa}
\end{equation}
shifting from exploration to exploitation as the budget depletes. On the same XAM slice (Fig.~\ref{fig:lcb}), Phase-B trials concentrate in the low-$J_b$ valley after Phase-A coverage.

\subsection{Pluggable ML Surrogates}
\label{sec:surrogate}
Phase~B treats every surrogate through the same interface: given training pairs $\mathcal{D}=\{(\theta_i,J_b(\theta_i))\}$, each refit returns $\mu(\theta)$ and $\sigma(\theta)$ for \eqref{eq:lcb}. RF and MLP aggregate $M$ members via
\begin{equation}
  \mu(\theta) = \tfrac{1}{M}\sum_{m=1}^{M}h_m(\theta),\quad
  \sigma(\theta) = \sqrt{\tfrac{1}{M}\sum_{m=1}^{M}\!\bigl(h_m(\theta)-\mu(\theta)\bigr)^{2}}.
  \label{eq:interface}
\end{equation}
The GP instead reads $\mu$ and $\sigma$ directly from the posterior \eqref{eq:gp_pred}.

\paragraph{Gaussian Process}
The GP~\cite{ref_pracbo} is a Bayesian non-parametric regressor with a Mat\'ern-$5/2$ kernel ($\nu{=}\tfrac{5}{2}$), signal scale $\sigma_f$, length scales $l_d$, and noise $\sigma_n$:
\begin{equation}
  \begin{aligned}
    k(\theta,\theta') &= \sigma_f^2\!\Bigl(1+\sqrt{5r}+\tfrac{5}{3}r\Bigr)\exp(-\sqrt{5r}),\\
    r &= \textstyle\sum_{d=1}^{10}(\theta_d-\theta'_d)^2/l_d^2.
  \end{aligned}
\end{equation}
Inputs are standardized and $J_b$ is $z$-scored before fitting; predictions are mapped back to the original scale. Given Gram matrix $K$ and cross-covariance $\mathbf{k}_*$, the posterior at $\theta_*$ is
\begin{equation}
  \begin{aligned}
    \mu(\theta_*) &= \mathbf{k}_*^{\top}(K+\sigma_n^2 I)^{-1}\mathbf{y},\\
    \sigma^2(\theta_*) &= k(\theta_*,\theta_*)-\mathbf{k}_*^{\top}(K+\sigma_n^2 I)^{-1}\mathbf{k}_*.
  \end{aligned}
  \label{eq:gp_pred}
\end{equation}
Hyperparameters maximize marginal likelihood at each refit.

\paragraph{Random Forest}
The RF regressor~\cite{ref_smac2011} uses $M{=}500$ bootstrap trees with max.\ depth~20 on raw $\theta$. Tree $m$ returns the leaf-mean prediction
\begin{equation}
  \begin{aligned}
    h_m^{\mathrm{RF}}(\theta) &= \sum_{\ell\in\mathcal{L}_m} c_{m\ell}\,\mathbf{1}\{\theta\in R_{m\ell}\}, \\
    c_{m\ell} &= \tfrac{1}{|\mathcal{I}_{m\ell}|}\sum_{i\in\mathcal{I}_{m\ell}} J_b(\theta_i),
  \end{aligned}
\end{equation}
where $\mathcal{L}_m$ is the leaf set of tree $m$, $R_{m\ell}$ are the leaf hyper-rectangles in $\Theta$, and $\mathcal{I}_{m\ell}$ indexes training points in leaf $\ell$. Inter-tree mean and spread supply $\mu$ and $\sigma$ via \eqref{eq:interface}.

\paragraph{MLP Ensemble}
The higher-capacity MLP ensemble~\cite{ref_ensemble} uses $M{=}10$ bootstrap-trained two-layer $(64,64)$ member networks, each optimized with limited-memory Broyden--Fletcher--Goldfarb--Shanno (L-BFGS). Inputs are min--max scaled to the per-scene bounds,
\begin{equation}
  \tilde{\theta}=(\theta-\theta_{\min})/(\theta_{\max}-\theta_{\min}),
\end{equation}
and member $m$ predicts
\begin{equation}
  h_m^{\mathrm{MLP}}(\theta)=\bar{J}+s_J\,\hat{y}_m(\tilde{\theta}),
\end{equation}
where $\hat{y}_m$ is the normalized network output and $\bar{J}$ and $s_J$ are the mean and std.\ of $J_b$ from the initial fit. Phase~B warm-starts member weights rather than refitting from scratch. Inter-member mean and spread supply $\mu$ and $\sigma$ via \eqref{eq:interface}.

\section{Experiments}
\label{sec:experiments}

\subsection{Setup}
\label{sec:setup}
\begin{figure}[h]
  \centering
  \subfloat[\subcap{Tianjin}]{\sceneimgpair{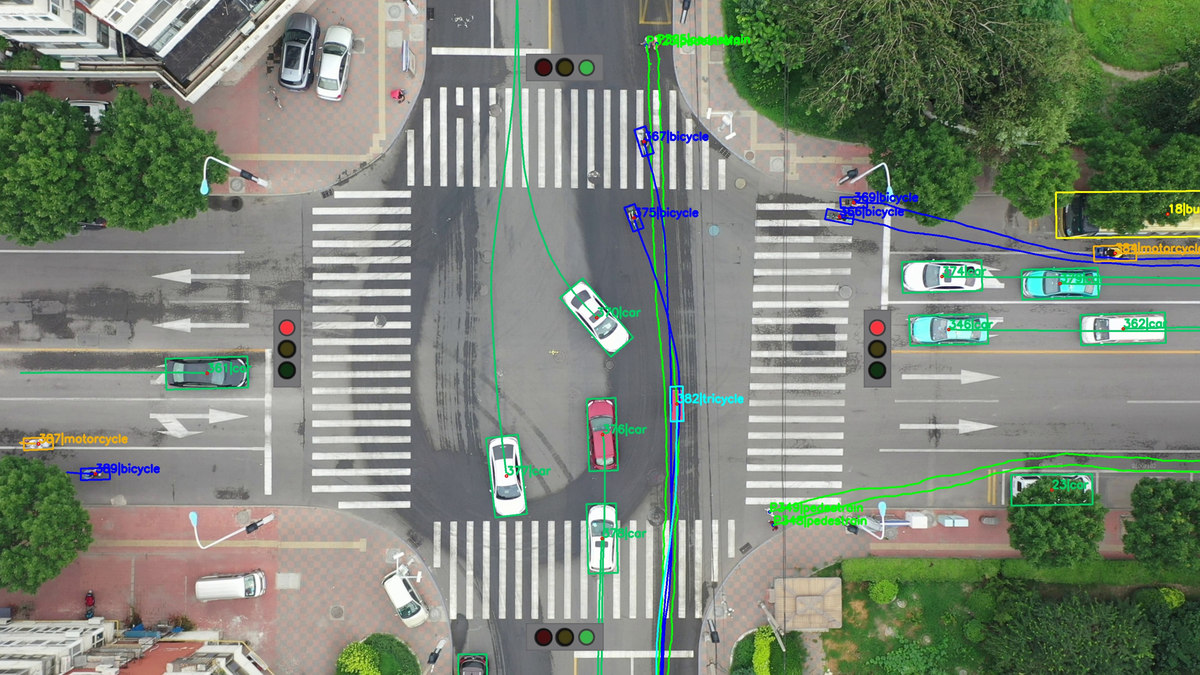}\label{fig:scene_tj}}
  \hfill
  \subfloat[\subcap{YTDJ}]{\sceneimgpair{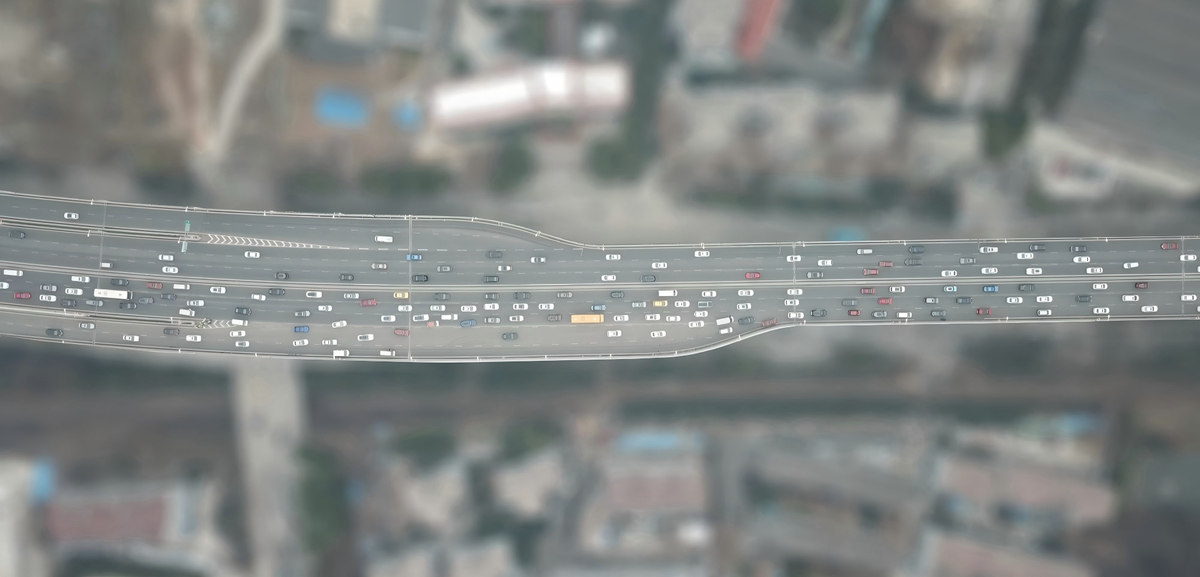}\label{fig:scene_ytdj}}
  \\
  \subfloat[\subcap{Changchun}]{\sceneimgpair{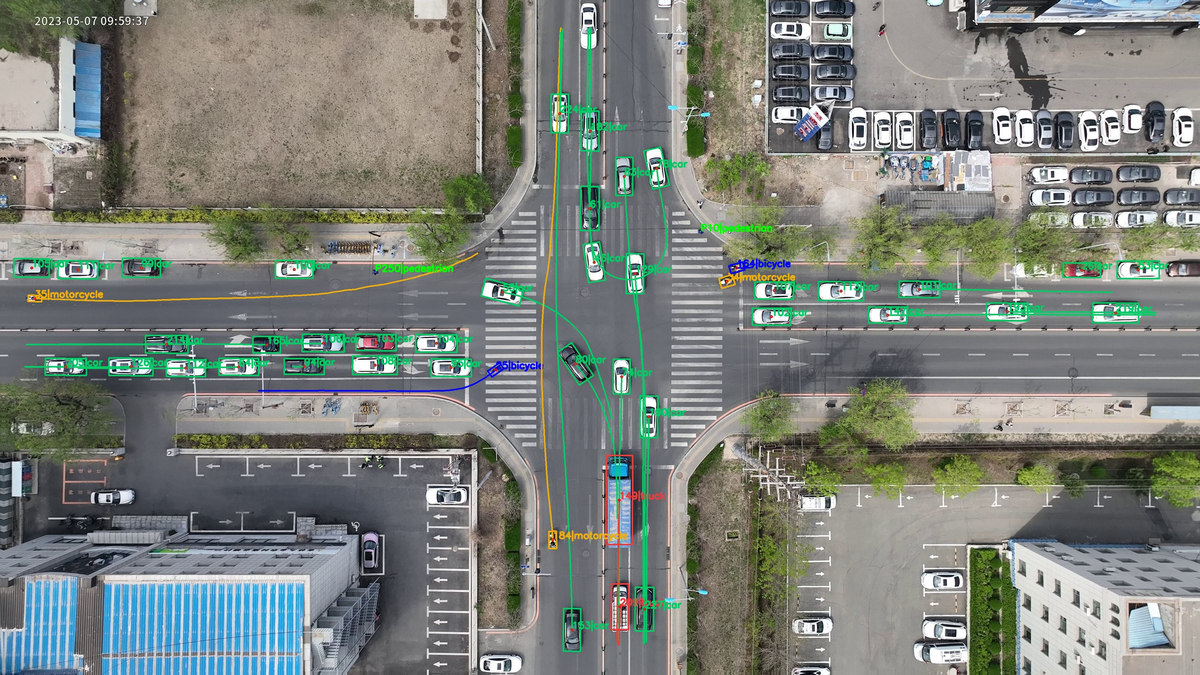}\label{fig:scene_cc}}
  \hfill
  \subfloat[\subcap{RML}]{\sceneimgpair{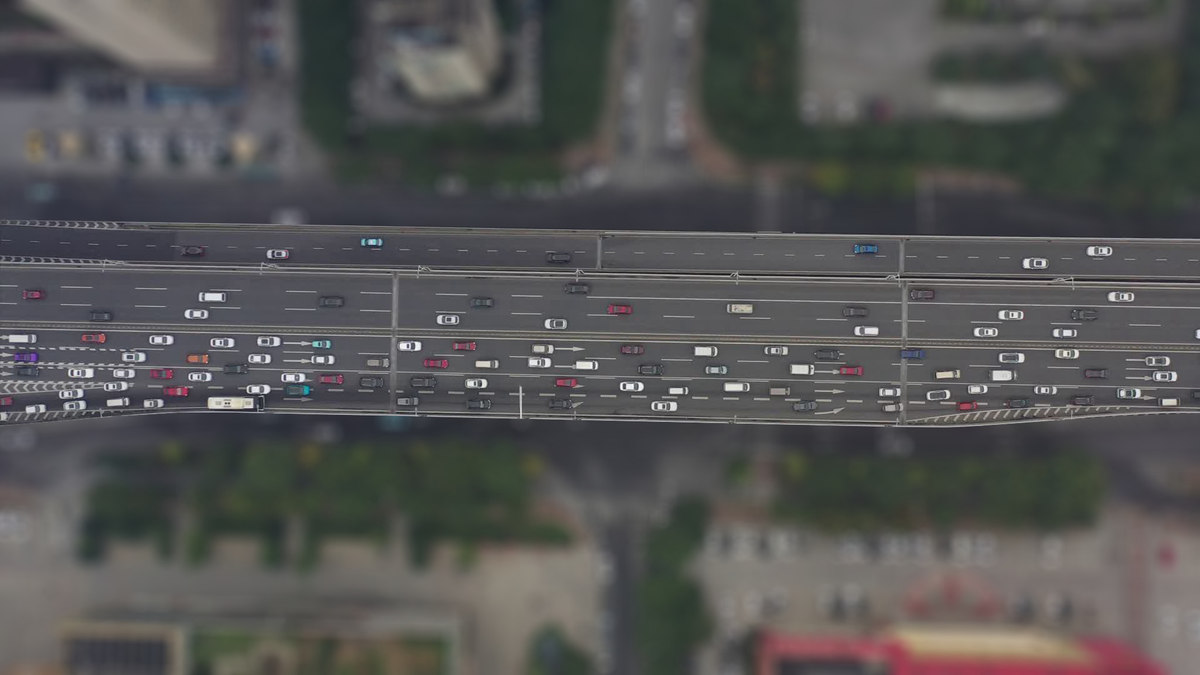}\label{fig:scene_rml}}
  \\
  \subfloat[\subcap{Xi'an}]{\sceneimgpair{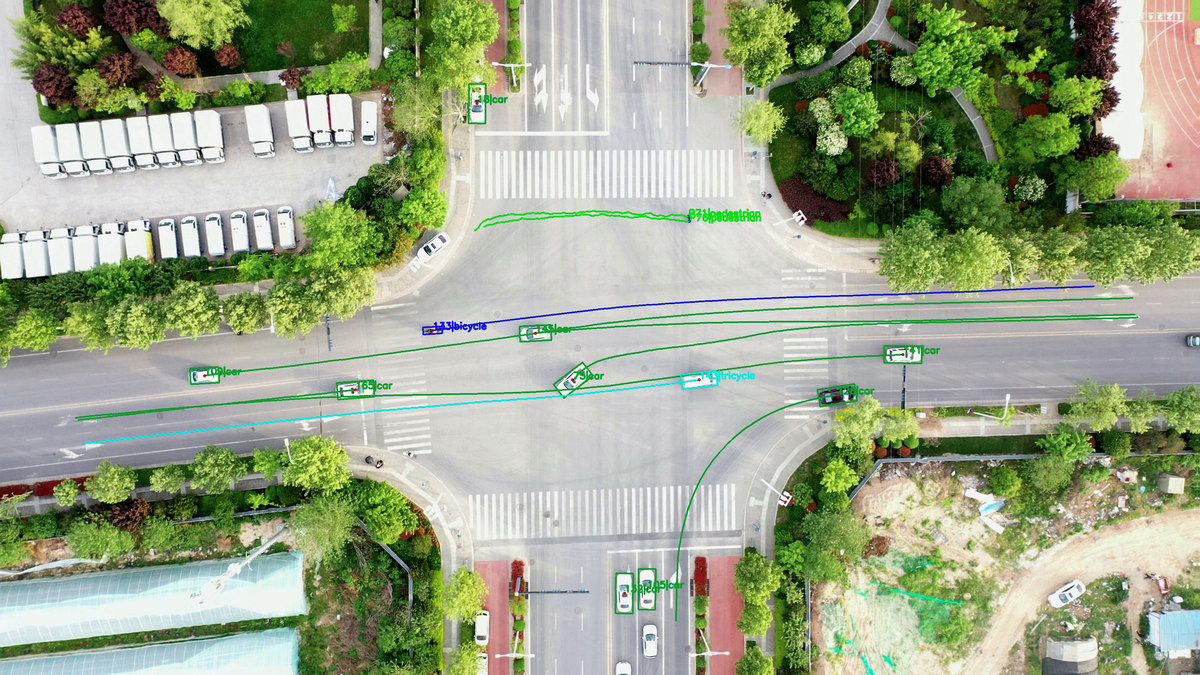}\label{fig:scene_xa}}
  \hfill
  \subfloat[\subcap{XAM}]{\sceneimgpair{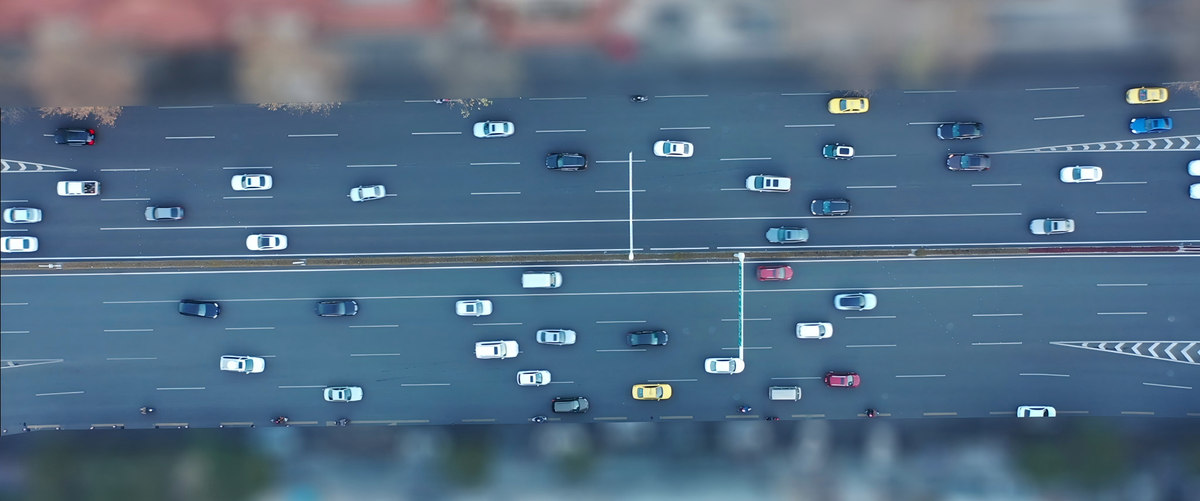}\label{fig:scene_xam}}
  \caption{Six heterogeneous evaluation scenes:
    \emph{Left:} SIND signalized intersections.
    \emph{Right:} UTE urban-expressway segments.}
  \label{fig:scenes}
\end{figure}

\textbf{Datasets.} We evaluate on six heterogeneous real-world scenes reconstructed as SUMO networks from SIND~\cite{ref_sind} and UTE~\cite{ref_ute} aerial trajectories (Fig.~\ref{fig:scenes}): signalized intersections (Tianjin, Changchun, Xi'an) spanning a low-to-high saturation range, and urban-expressway segments (YTDJ, RML, XAM) covering merge and weaving bottlenecks.

\textbf{Budget.} Recalibration shares each rolling update window with data processing and prediction~\cite{ref_huainan2022}. With serial SUMO--TraCI replays taking seconds, we use a representative calibration budget of $B{=}100$ simulations per scene.

\textbf{Protocol.} Within each scene, all methods use identical parameter bounds and share $B{=}100$, objective $J_b$, IVW weights, and seeds $\{42,101,202,303,404\}$, with a fresh SUMO seed per simulation run. FLAT uses $N_{\mathrm{init}}{=}40$ initial LHS samples and an $18\%$ trust region. We report best-so-far $J_b$.

\textbf{Baselines.} We instantiate FLAT with three interchangeable ML surrogates, GP, RF, and MLP, and compare against four optimizer families: SPSA~\cite{ref_spsa} (gradient estimation), GA (population search), CMA-ES~\cite{ref_cmaes} (evolution strategy), and TPE~\cite{ref_tpe} (density-estimation Bayesian optimizer).\ifdefined\FLATarxiv{} Appendix~\ref{app:protocols} details the optimizer settings and serial evaluation costs.\fi

\textbf{Significance.} We average the five runs within each scene and apply two-sided exact paired Wilcoxon tests~\cite{ref_wilcoxon} across the six scene-level observations. Statistical evidence is reported jointly with effect sizes and cross-scene win counts.

\subsection{Best Under Equal Budget}
\label{sec:results}
Under equal budget, FLAT attains the lowest six-scene mean $J_b$ (Table~\ref{tab:main}). GP reaches $0.231$, with RF and MLP at $0.241$ and $0.240$, well ahead of TPE ($+11.2\%$ vs.\ GP), GA ($+19.6\%$), CMA-ES ($+21.2\%$), and SPSA ($+66.0\%$). Because the four baselines span distinct optimizer mechanisms, the margin is unlikely to reflect a coincidental match to one class.
\begin{table}[h]
  \caption{Six-scene mean $J_b$ at budget $100$ (mean\,$\pm$\,seed std over five seeds). $\Delta\%$: relative to each FLAT variant's six-scene mean. Wins/6: FLAT-GP scene wins; $p$: scene-level test.}
  \label{tab:main}
  \centering\scriptsize\setlength\tabcolsep{2pt}\renewcommand{\arraystretch}{1.1}
  \begin{tabularx}{\columnwidth}{@{}>{\columncolor{white}[0pt][\tabcolsep]}L c c c c c >{\columncolor{white}[\tabcolsep][0pt]}c@{}}
    \toprule
    Method & \mbox{Mean $J_b\downarrow$} & vs.\ GP & vs.\ RF & vs.\ MLP & Wins/6 & $p$ \\
    \specialrule{\lightrulewidth}{\aboverulesep}{0pt}
    \rowcolor{black!5}
\rule{0pt}{\dimexpr1.1\ht\strutbox+\belowrulesep\relax}FLAT-GP & \rule{0pt}{\dimexpr1.1\ht\strutbox+\belowrulesep\relax}\textbf{0.231}\,\std{0.003} & \rule{0pt}{\dimexpr1.1\ht\strutbox+\belowrulesep\relax}--- & \rule{0pt}{\dimexpr1.1\ht\strutbox+\belowrulesep\relax}$-4.0$ & \rule{0pt}{\dimexpr1.1\ht\strutbox+\belowrulesep\relax}$-3.9$ & \rule{0pt}{\dimexpr1.1\ht\strutbox+\belowrulesep\relax}--- & \rule{0pt}{\dimexpr1.1\ht\strutbox+\belowrulesep\relax}---    \\
    \rowcolor{black!5}
    FLAT-RF  & 0.241\,\std{0.009}          & $+4.2$ & ---    & $+0.1$ & ---    &  ---    \\
    \rowcolor{black!5}
    FLAT-MLP & 0.240\,\std{0.006}          & $+4.1$ & $-0.1$ & ---    & ---    &  ---    \\
    TPE      & 0.257\,\std{0.010}          & $+11.2$ & $+6.7$ & $+6.8$ & 5/6 & 0.0625 \\
    GA       & 0.276\,\std{0.013}          & $+19.6$ & $+14.7$ & $+14.9$ & 6/6 & 0.0313 \\
    CMA-ES   & 0.280\,\std{0.023}          & $+21.2$ & $+16.3$ & $+16.4$ & 6/6 & 0.0313 \\
    SPSA     & 0.383\,\std{0.043}          & $+66.0$ & $+59.3$ & $+59.5$ & 6/6 & 0.0313 \\
    \bottomrule
  
  \end{tabularx}
\end{table}

\begin{figure}[h]
  \centering
  \plotpanelfull{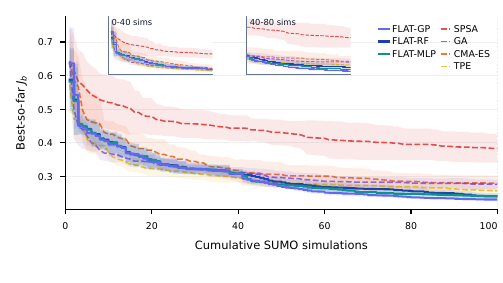}
  \caption{Running best-so-far $J_b$ for seven optimizers, averaged over six scenes and five seeds. \emph{Shaded bands:} $\pm1$\,std across seeds. \emph{Dotted line:} budget cap $B{=}100$. \emph{Insets:} zoomed 0--40 and 40--80 simulations.}
  \label{fig:panel}
\end{figure}

\begin{figure}[h]
  \centering
  \plotpanelfull{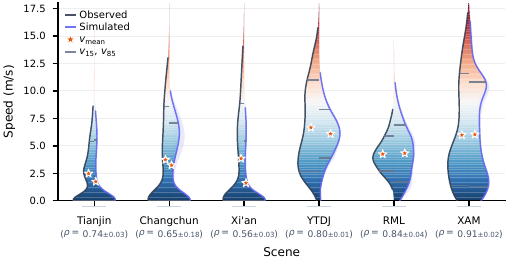}
  \caption{Observed vs.\ FLAT-GP simulated speed distributions (five seeds); simulated mean kernel density estimation (KDE) $\pm1$\,std envelope. Per-scene $\rho{=}1{-}D_{\mathrm{KS}}$ (mean$\,\pm$\,seed std); stars/ticks: $v_{\mathrm{mean}}$, $v_{15}$, $v_{85}$.}
  \label{fig:speeddist}
\end{figure}

Fig.~\ref{fig:panel} shows continued gains after the initial LHS coverage: from simulation runs $40$ to $100$, FLAT reduces mean $J_b$ by $22$--$25\%$ across all three surrogates, versus $11$--$14\%$ for the four baselines. GP also has the narrowest seed spread ($\pm0.003$, versus $\pm0.010$--$0.043$ for the baselines). Fig.~\ref{fig:speeddist} cross-checks $J_b$ against observed speed distributions (similarity $\bar{\rho}{=}0.75$), with expressways closest ($\rho{\geq}0.80$, tight seed bands). Aligned $v_{15}$, $v_{\mathrm{mean}}$, and $v_{85}$ markers confirm that the simulations capture both the mean and spread of observed speeds.

\subsection{Cheapest to Equal Quality}
\label{sec:efficiency}
Equal-budget error (\S\ref{sec:results}) does not establish whether baselines close the gap with more simulations. We extend each to $B{=}500$ and record, per scene$\times$seed pair, the first run whose running-best $J_b$ meets FLAT's per-scene mean at $B{=}100$ (Table~\ref{tab:efficiency}). TPE's mean evaluation count is $1.47\times$--$2.21\times$ FLAT's budget ($\pm 139$--$162$ simulations across pairs), while CMA-ES and GA average $2.76\times$--$4.14\times$. Furthermore, only $28/30$ pairs for TPE and $9/30$ for SPSA reach the FLAT-RF threshold within the $500$-sim cap; the gap is a failure to match, not just slower convergence. This sample efficiency is valuable under tight recalibration budgets.

\begin{table}[h]
  \caption{Evaluation counts to match FLAT@100, mean\,$\pm$\,std over $30$ scene\,$\times$\,seed pairs; runs not reaching the threshold within $500$ evaluations are assigned $501$. Reach: pairs meeting the FLAT-RF threshold within budget.}
  \label{tab:efficiency}
  \centering\scriptsize\setlength\tabcolsep{2pt}\renewcommand{\arraystretch}{1.1}
  \begin{tabularx}{\columnwidth}{@{}l C C C C C C C@{}}
    \toprule
    & \multicolumn{2}{c}{vs.\ GP} & \multicolumn{2}{c}{vs.\ RF} & \multicolumn{2}{c}{vs.\ MLP} & \\
    \cmidrule(lr){2-3}\cmidrule(lr){4-5}\cmidrule(lr){6-7}
    Baseline & Sims$\downarrow$ & Ratio & Sims$\downarrow$ & Ratio & Sims$\downarrow$ & Ratio & Reach$\uparrow$ \\
    \midrule
    TPE    & 221\,\std{162} & 2.21$\times$ & 167\,\std{139} & 1.67$\times$ & 147\,\std{140} & 1.47$\times$ & 28/30 \\
    CMA-ES & 402\,\std{107} & 4.02$\times$ & 288\,\std{148} & 2.88$\times$ & 276\,\std{108} & 2.76$\times$ & 25/30 \\
    GA     & 414\,\std{142} & 4.14$\times$ & 360\,\std{165} & 3.60$\times$ & 305\,\std{171} & 3.05$\times$ & 15/30 \\
    SPSA   & 442\,\std{131} & 4.42$\times$ & 409\,\std{159} & 4.09$\times$ & 424\,\std{154} & 4.24$\times$ &  9/30 \\
    \bottomrule
  \end{tabularx}
\end{table}

\begin{figure*}[!t]
\begin{minipage}[t]{0.49\textwidth}
\centering
  \subfloat[\subcap{Tianjin}]{\plotpanelpair{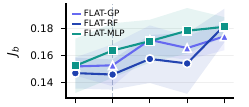}\label{fig:ninit_tj}}
  \hfill
  \subfloat[\subcap{YTDJ}]{\plotpanelpair{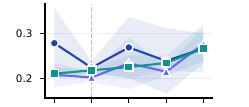}\label{fig:ninit_ytdj}}
  \\[-1pt]
  \subfloat[\subcap{Changchun}]{\plotpanelpair{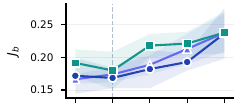}\label{fig:ninit_cc}}
  \hfill
  \subfloat[\subcap{RML}]{\plotpanelpair{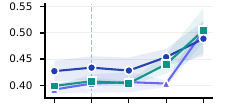}\label{fig:ninit_rml}}
  \\[-1pt]
  \subfloat[\subcap{Xi'an}]{\plotpanelpair{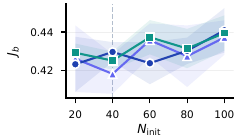}\label{fig:ninit_xa}}
  \hfill
  \subfloat[\subcap{XAM}]{\plotpanelpair{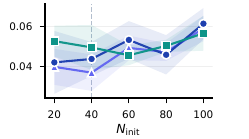}\label{fig:ninit_xam}}
  \caption{$N_{\mathrm{init}}$ ablation: final $J_b$ vs.\ warm-up size at fixed budget $100$, five-seed mean $\pm1$\,std. $N_{\mathrm{init}}{=}100$ reduces to pure LHS.}
  \label{fig:ninit}
\end{minipage}\hfill
\begin{minipage}[t]{0.49\textwidth}
\centering
  \subfloat[\subcap{Tianjin}]{\plotpanelpair{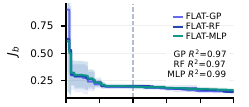}\label{fig:calib_tj}}
  \hfill
  \subfloat[\subcap{YTDJ}]{\plotpanelpair{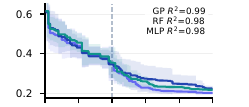}\label{fig:calib_ytdj}}
  \\[-1pt]
  \subfloat[\subcap{Changchun}]{\plotpanelpair{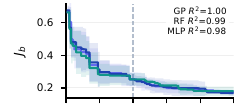}\label{fig:calib_cc}}
  \hfill
  \subfloat[\subcap{RML}]{\plotpanelpair{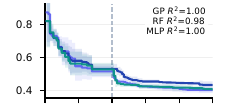}\label{fig:calib_rml}}
  \\[-1pt]
  \subfloat[\subcap{Xi'an}]{\plotpanelpair{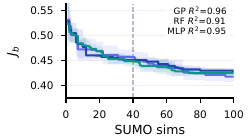}\label{fig:calib_xa}}
  \hfill
  \subfloat[\subcap{XAM}]{\plotpanelpair{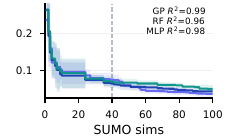}\label{fig:calib_xam}}
  \caption{Per-scene best-so-far $J_b$ convergence for three surrogates, five-seed mean $\pm1$\,std. $R^2$ annotations report surrogate fit at the final training set.}
  \label{fig:percalib}
\end{minipage}
\end{figure*}

\begin{table}[!t]
  \caption{Objective ablation: final $J_b$ when training on $J_b$ vs.\ $J_s$ (always scored by $J_b$). Five-seed mean\,$\pm$\,std; 6/6 scenes, scene-level exact Wilcoxon $p{=}0.031$ per optimizer.}
  \label{tab:objabl}
  \centering\scriptsize\setlength\tabcolsep{2.5pt}\renewcommand{\arraystretch}{1.1}
  \begin{tabularx}{\columnwidth}{@{}l CCCC@{}}
    \toprule
    Scene & FLAT-RF ($J_b$) & FLAT-RF ($J_s$) & TPE ($J_b$) & TPE ($J_s$) \\
    \midrule
    Tianjin   & \textbf{0.146}\,\std{0.011} & 0.345\,\std{0.066} & \textbf{0.165}\,\std{0.010} & 0.327\,\std{0.116} \\
    Changchun & \textbf{0.168}\,\std{0.016} & 0.279\,\std{0.041} & \textbf{0.167}\,\std{0.013} & 0.277\,\std{0.068} \\
    Xi'an     & \textbf{0.430}\,\std{0.005} & 0.488\,\std{0.014} & \textbf{0.426}\,\std{0.013} & 0.471\,\std{0.017} \\
    YTDJ      & \textbf{0.223}\,\std{0.014} & 0.467\,\std{0.129} & \textbf{0.284}\,\std{0.067} & 0.435\,\std{0.102} \\
    RML       & \textbf{0.434}\,\std{0.022} & 0.514\,\std{0.016} & \textbf{0.452}\,\std{0.008} & 0.620\,\std{0.024} \\
    XAM       & \textbf{0.044}\,\std{0.008} & 0.105\,\std{0.097} & \textbf{0.047}\,\std{0.008} & 0.234\,\std{0.107} \\
    \bottomrule
  \end{tabularx}
\end{table}

\subsection{Sources of Performance Gain}
\label{sec:gain}
We examine the roles of objective design, sequential acquisition, and surrogate choice through three ablations.

\textbf{Objective geometry dominates.} Holding the optimizer and budget fixed, we train on $J_s$ or $J_b$ but always score $\theta^\star$ by $J_b$. Table~\ref{tab:objabl} shows that training on $J_s$ degrades FLAT-RF and TPE in 6/6 scenes (scene-level exact Wilcoxon $p{=}0.031$ each). Final $J_b$ rises by $80.6\%$ on average under FLAT-RF and more than doubles on Tianjin, XAM, and YTDJ. Surrogates still fit $J_s$, yet the returned parameters miss behavioral alignment; the same degradation under TPE rules out an acquisition-specific artifact. $J_b$ alone occupies the middle rung between \eqref{eq:traj} and \eqref{eq:js} (Fig.~\ref{fig:landscape}).
Appendix~\ref{app:geometry} traces this mechanism through landscape smoothness, held-out prediction, and uncertainty.

\textbf{Sequential acquisition contributes.} Holding $J_b$ and $B{=}100$ fixed, we sweep $N_{\mathrm{init}}$ (Fig.~\ref{fig:ninit}); $N_{\mathrm{init}}{=}100$ reduces Algorithm~\ref{alg:flat} to Phase~A. With GP, the six-scene mean stays near $0.23$ for $N_{\mathrm{init}}{=}20$--$40$ but rises to $0.28$ at $100$, $21.7\%$ worse than at $40$, with pure LHS worst or near-worst (e.g.\ RML $0.403{\to}0.503$). Rugged scenes account for most of the penalty, whereas Xi'an shifts only $5\%$ across the sweep, consistent with a fingerprint floor leaving little refinement headroom. $N_{\mathrm{init}}{=}40$ gives a comparable mean to $20$, cuts seed variance on YTDJ (std $0.010$ vs.\ $0.028$), and reserves $60$ evaluations for LCB.

\textbf{Surrogate choice is secondary.} With $J_b$ and acquisition fixed, GP, RF, and MLP track closely on average (Fig.~\ref{fig:percalib}). On XAM, Tianjin, and Changchun, all three finish within $0.02$. GP has the lowest global mean and tightest seed spread (Table~\ref{tab:main}), fitting an uncertainty-aware LCB backend; on harder scenes (Xi'an, YTDJ, RML) MLP gains most, especially RML ($J_b$ $0.434{\to}0.408$, $R^2$ $0.976{\to}0.997$). Replaying fixed final parameters from four matched repeats on ten shared SUMO seeds not used during optimization gives a maximum gap of $4.1\%$ in six-scene mean $J_b$. No surrogate dominates; backend differences remain small relative to the $80.6\%$ increase in behavioral error under objective ablation.

\subsection{Operating Range}
\label{sec:limits}
Sample-efficient guidance works where $J_b$ is learnable yet baselines waste evaluations (YTDJ, RML, Changchun). XAM ($J_b\approx0.044$) sets the easy bound, with FLAT, TPE, and GA within $0.01$. Xi'an sets the hard bound at our SIND trio's high-saturation end, where signal cycling yields a stop-go fingerprint shared $\theta$ cannot match. Fig.~\ref{fig:speeddist} gives $\rho{=}0.56$ with speeds biased low relative to post-queue acceleration, a scene-wide instance of the fit trade-offs Punzo~\textit{et~al.}~\cite{ref_punzo2021} establish pairwise. Every optimizer family clusters at $J_b\approx0.43$, SPSA included.

The common plateau may reflect a limitation of the Krauss car-following model. The Intelligent Driver Model (IDM) and psychophysical Wiedemann models represent stop-and-go transients differently. We keep Krauss and the parameterization fixed under one TraCI protocol to isolate fingerprint smoothing and LCB rather than car-following choice. Practical gains lie between XAM's basin and Xi'an's ceiling for scenes rugged enough to reward guidance yet below Xi'an's model limit.

\section{Conclusion}
\label{sec:conclusion}

We presented FLAT, a framework that addresses a central bottleneck in surrogate-assisted traffic calibration: raw trajectory objectives can be too rugged to learn from few samples, leaving acquisition without a reliable signal. A behavioral fingerprint smooths the parameter--error landscape so a surrogate can fit it from a few dozen samples, and annealed LCB acquisition concentrates the budget where it most reduces error.

Across six heterogeneous scenes and four optimizer families, FLAT achieves the lowest scene-averaged error under matched budgets; even with up to 500 simulations, competitors need 1.5 to 4.4 times more simulations to match FLAT's $100$-run quality. Ablations confirm that the gain traces to objective geometry and sequential allocation rather than surrogate capacity: training on $J_s$ instead of $J_b$ raises final behavioral error by $80.6\%$ on average, pure space-filling raises the six-scene mean by $21.7\%$, and the three surrogates differ by at most $4.2\%$ in global mean. These results highlight objective geometry as a route to sample-efficient black-box optimization.

Future work will extend this insight through self-supervised encoders pretrained on corpora such as Waymo and nuPlan, deriving surrogate-learnable objectives offline and reserving scarce simulations for sequential refresh. Rolling digital twins at network scale should compose warm-started scene calibrations across observation windows, spending those budgets where mismatch drifts fastest. Transfer to other urban-mobility models and regional climate emulation remains to be tested.

\appendices

\section{Smoothness, Learnability and Uncertainty}
\label{app:geometry}

This appendix traces whether slice-level smoothing persists in 10-D $\Theta$, whether FLAT-GP can learn it at our calibration budget, and how its posterior fits held-out data. We report all three on XAM and Xi'an, our easy and hard bounds.

\subsection{From Steepness to Smoothness}
\label{sec:apx_geometry}
\begin{figure}[h]
  \centering
  \subfloat[\subcap{XAM}]{\plotpanelcomposite{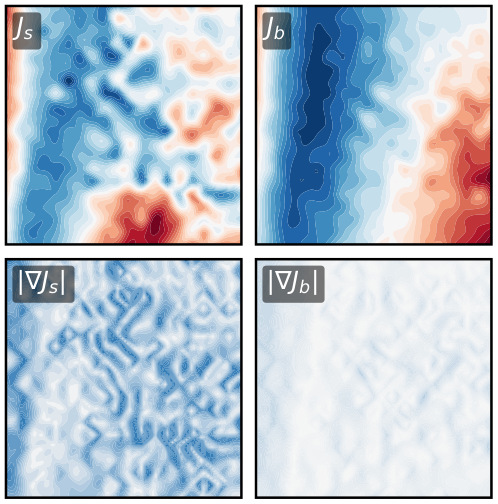}\label{fig:apxcomp_xam}}
  \hfill
  \subfloat[\subcap{Xi'an}]{\plotpanelcomposite{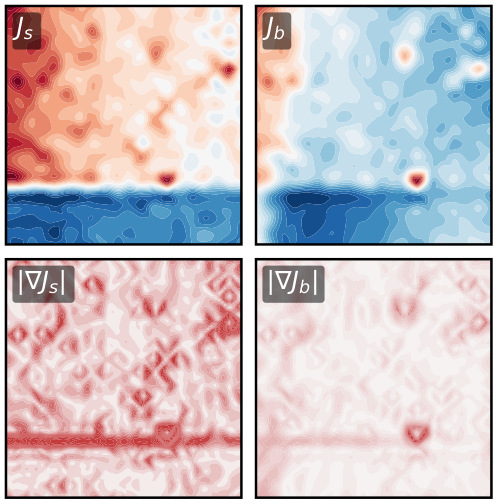}\label{fig:apxcomp_xian}}
  \caption{Objective landscapes in the $(a_{\max},\tau)$ plane on XAM and Xi'an.
    \emph{Top row:} $J_s$ and $J_b$. \emph{Bottom row:} $|\nabla J_s|$ and $|\nabla J_b|$.}
  \label{fig:apxcomposite}
\end{figure}

Fig.~\ref{fig:apxcomposite} extends Fig.~\ref{fig:landscape} in the same $(a_{\max},\tau)$ window, contrasting $J_s$ and $J_b$ with $|\nabla J|$. On XAM, $J_b$ forms one wide, connected low-error basin with sparse gradient peaks, while $J_s$ fragments into narrower, steeper cells. Xi'an is rougher for both objectives, yet $J_b$ retains a single dominant basin where $J_s$ does not. A sharp low-$\tau$ ridge marks queue discharge, where tighter headway triggers stop-and-go and a fingerprint jump.

\begin{figure}[h]
  \centering
  \subfloat[\subcap{XAM}]{\plotpanelpair{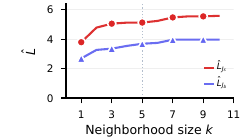}\label{fig:apxlip_xam}}
  \hfill
  \subfloat[\subcap{Xi'an}]{\plotpanelpair{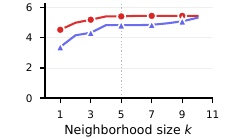}\label{fig:apxlip_xian}}
  \caption{Standardized local Lipschitz proxy $\hat L$ \eqref{eq:lipschitz} vs.\ neighborhood size $k$ on a 10-D $400$-point LHS design; dotted line marks the reported $k{=}5$.}
  \label{fig:apxlipschitz}
\end{figure}

To test this contrast beyond the 2-D slice, we estimate a \textit{local Lipschitz proxy} on a $400$-point LHS over 10-D $\theta$. We scale each objective by its standard deviation $s_J$ over this design, then take $k$ nearest neighbors of each $\theta_i$ in the same min--max normalized $\tilde\theta$ as \S\ref{sec:surrogate} and record the steepest slope,
\begin{equation}
  \hat{L}_k(\theta_i) = \max_{j\in\mathcal{N}_k(\theta_i)} \frac{|J(\theta_i)-J(\theta_j)|}{s_J\,\|\tilde\theta_i-\tilde\theta_j\|_2},
  \label{eq:lipschitz}
\end{equation}
with per-scene $\hat L = \mathrm{p95}_i\{\hat L_k(\theta_i)\}$, robust to occasional SUMO replay noise. Fig.~\ref{fig:apxlipschitz} shows that $J_b$ retains lower standardized slopes at every $k\in\{1,\ldots,10\}$ on both scenes, even against the vehicle-averaged $J_s$ in \eqref{eq:js}. At $k{=}5$, the standardized slope ratio $\hat L_{J_s}/\hat L_{J_b}$ is $1.39$ on XAM and $1.12$ on Xi'an.

\subsection{From Smoothness to Learnability}
\label{sec:apx_learnability}
\begin{figure}[h]
  \centering
  \subfloat[\subcap{XAM}]{\plotpanelpair{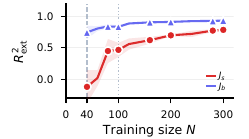}\label{fig:apxr2_xam}}
  \hfill
  \subfloat[\subcap{Xi'an}]{\plotpanelpair{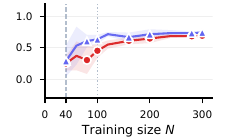}\label{fig:apxr2_xian}}
  \caption{Held-out $R^2_{\mathrm{ext}}$ vs.\ training size $N$ for FLAT-GP fit on $J_s$ vs.\ $J_b$; dotted line marks $N{=}100$, the calibration budget.}
  \label{fig:apxr2}
\end{figure}

Smoothness alone does not guarantee that a surrogate can learn the objective from few samples. From the same $400$-point 10-D pool, we fix a $100$-point holdout for both objectives and all training sizes, and fit FLAT-GP on five random $N$-point subsets of the remaining $300$ (Fig.~\ref{fig:apxr2}).

At $N{=}100$, held-out $R^2_{\mathrm{ext}}$ is $0.834$ for $J_b$ versus $0.467$ for $J_s$ on XAM, and $0.634$ versus $0.453$ on Xi'an. Fig.~\ref{fig:percalib} reports in-sample $R^2$ on sequential data. $J_b$ thus pairs the behavioral gains in Table~\ref{tab:objabl} with better held-out prediction.

\subsection{From Learnability to Uncertainty}
\label{sec:apx_calib}
On the same holdout as \S\ref{sec:apx_learnability}, we jointly assess GP mean and uncertainty $(\mu,\sigma)$ at $N{=}100$ with negative log-likelihood (NLL) and continuous ranked probability score (CRPS). Scores use the training-response $z$-score scale (\S\ref{sec:surrogate}):
\begin{align}
  \mathrm{NLL} &= \mathbb{E}\!\left[\tfrac{1}{2}\log(2\pi\sigma^2)+\tfrac{(y-\mu)^2}{2\sigma^2}\right], \label{eq:apx_nll}\\
  \mathrm{CRPS} &= \mathbb{E}\!\int\!\left(F(z)-\mathbf{1}\{z\ge y\}\right)^{2}\mathrm{d}z,\quad F=\Phi\!\left(\tfrac{\cdot-\mu}{\sigma}\right), \label{eq:apx_crps}
\end{align}
As a reference, we use a constant Gaussian with each training subset's response mean and variance, equivalent to $\mathcal N(0,1)$ on this scale. For training subset $r$, CRPS gain is
\begin{equation}
  G_r = 100\bigl(1-\mathrm{CRPS}_{\mathrm{GP},r}/\mathrm{CRPS}_{\mathrm{ref},r}\bigr).
  \label{eq:apx_gain}
\end{equation}

Table~\ref{tab:apxcalib} extends the learnability advantage to probabilistic prediction: compared with $J_s$, $J_b$ gives lower mean NLL on both scenes, while mean CRPS falls by $43\%$ on the XAM expressway and remains comparable at the Xi'an intersection. GP outperforms the reference for both objectives, with the largest CRPS gain when modeling $J_b$ on XAM.

\begin{table}[H]
  \caption{Standardized GP predictive scores on held-out landscapes ($N{=}100$); five-subsample mean\,$\pm$\,std.}
  \label{tab:apxcalib}
  \centering\scriptsize\setlength\tabcolsep{3.5pt}\renewcommand{\arraystretch}{1.15}
  \begin{tabular*}{\columnwidth}{@{\extracolsep{\fill}}lcccc@{}}
    \toprule
    & \multicolumn{2}{c}{XAM} & \multicolumn{2}{c}{Xi'an} \\
    \cmidrule(lr){2-3}\cmidrule(lr){4-5}
    & FLAT-GP ($J_b$) & FLAT-GP ($J_s$) & FLAT-GP ($J_b$) & FLAT-GP ($J_s$) \\
    \midrule
    NLL $\downarrow$ & \tjb{0.541}\,\std{0.143} & 1.046\,\std{0.109} & \tjb{1.011}\,\std{0.278} & 1.308\,\std{0.418} \\
    CRPS $\downarrow$ & \tjb{0.211}\,\std{0.017} & 0.372\,\std{0.033} & \tjb{0.347}\,\std{0.071} & 0.359\,\std{0.058} \\
    CRPS gain (\%) $\uparrow$ & \tjb{62.5}\,\std{2.8} & 29.2\,\std{6.4} & \tjb{34.3}\,\std{6.2} & 30.3\,\std{8.9} \\
    \bottomrule
  \end{tabular*}
\end{table}

Together, the landscape, learnability, and uncertainty analyses support FLAT's sample-efficient calibration.

\section{Calibrated Behavioral Parameters}
\label{app:params}

This appendix gives physical meanings and the search box $\Theta$ in Table~\ref{tab:apxparams} with a priori bounds that bracket plausible Krauss car-following and LC2013 lane-changing behavior across scenes, fixed before optimization for all methods.

\begin{table}[h]
  \caption{TraCI keys, physical meanings, bounds, and units for six car-following and four lane-changing parameters.}
  \label{tab:apxparams}
  \centering\scriptsize\setlength\tabcolsep{2.5pt}\renewcommand{\arraystretch}{1.15}
  \begin{tabularx}{\columnwidth}{@{}l @{\hspace{30pt}} >{\raggedright\arraybackslash}X c c@{}}
    \toprule
    TraCI key & Physical meaning & Bounds & Unit \\
    \midrule
    \multicolumn{4}{@{}l}{\textit{Car-following (Krauss)}} \\
    \addlinespace[1pt]
    \texttt{accel} & Maximum acceleration $a_{\max}$ & $[1.0,\,3.5]$ & m/s$^2$ \\
    \texttt{decel} & Comfortable deceleration & $[2.0,\,6.0]$ & m/s$^2$ \\
    \texttt{sigma} & Driver imperfection & $[0.1,\,0.9]$ & --- \\
    \texttt{tau} & Desired time headway $\tau$ & $[0.5,\,2.5]$ & s \\
    \texttt{minGap} & Minimum standing gap & $[1.0,\,4.0]$ & m \\
    \texttt{speedFactor} & Desired-speed multiplier & $[0.5,\,1.2]$ & --- \\
    \midrule
    \multicolumn{4}{@{}l}{\textit{Lane-changing (LC2013)}} \\
    \addlinespace[1pt]
    \texttt{lcStrategic} & Strategic/mandatory eagerness & $[0.35,\,2.80]$ & --- \\
    \texttt{lcCooperative} & Cooperative willingness & $[0.35,\,2.00]$ & --- \\
    \texttt{lcAssertive} & Lane-changing assertiveness & $[0.35,\,2.80]$ & --- \\
    \texttt{lcSpeedGain} & Discretionary speed-gain weight & $[0.35,\,2.80]$ & --- \\
    \bottomrule
  \end{tabularx}
\end{table}

\ifdefined\FLATarxiv
  \section{Optimizer Protocols and Computational Cost}
\label{app:protocols}

Table~\ref{tab:protocols} specifies optimizer settings for the shared protocol in \S\ref{sec:setup}. All simulator calls count toward $B{=}100$, including initialization and both calls in each SPSA perturbation pair. The extended searches in \S\ref{sec:efficiency} retain these baseline settings with a cap of 500 simulator evaluations per run.

\begin{table}[h]
  \caption{Optimizer settings for the equal-budget comparison.}
  \label{tab:protocols}
  \centering\footnotesize
  \setlength\tabcolsep{3pt}\renewcommand{\arraystretch}{1.15}
  \begin{tabularx}{\columnwidth}{@{}l X@{}}
    \toprule
    Method & Configuration \\
    \midrule
    FLAT & 40-point LHS + 60 sequential LCB evaluations; $\kappa$ decreases linearly from 2.0 to 0.5; 2,500 local and 2,500 global candidates per step; local half-width 18\% of each parameter span; minimum span-normalized Euclidean distance 0.02 to evaluated points. \\
    SPSA & Random start; $a{=}0.22$, $c{=}0.15$, $A{=}20$, $\alpha{=}0.602$, $\gamma{=}0.101$; perturbations and updates in original coordinates, with componentwise clipping to the parameter bounds. \\
    GA & Random population of 10; tournament size 3; uniform crossover; Gaussian mutation with SD of 5\% of parameter span. \\
    CMA-ES & Random start in normalized $[0,1]$ coordinates; $\sigma_0{=}0.3$; bounded search with implementation-default adaptation. \\
    TPE & Seeded Optuna \texttt{TPESampler} with implementation-default settings; one simulator evaluation per trial. \\
    \bottomrule
  \end{tabularx}
\end{table}

\textbf{Evaluation cost.} On an Intel Core i7-11800H laptop running Windows~11 and SUMO~1.22.0, we timed three serial replays per scene at the FLAT-GP final parameters from run~101, using shared SUMO seeds $\{20001,20002,20003\}$. Across these 18 evaluations, scene-mean times ranged from $2.23$\,s on RML to $10.54$\,s on YTDJ. The timer includes SUMO startup, TraCI collection, fingerprint extraction and scoring, but excludes reference preprocessing and surrogate fitting.

\fi

\ifdefined\FLATrefreshbib
  \bibliographystyle{IEEEtran}
  \bibliography{root}

\begin{thebibliography}{10}
\providecommand{\url}[1]{#1}
\csname url@samestyle\endcsname
\providecommand{\newblock}{\relax}
\providecommand{\bibinfo}[2]{#2}
\providecommand{\BIBentrySTDinterwordspacing}{\spaceskip=0pt\relax}
\providecommand{\BIBentryALTinterwordstretchfactor}{4}
\providecommand{\BIBentryALTinterwordspacing}{\spaceskip=\fontdimen2\font plus
\BIBentryALTinterwordstretchfactor\fontdimen3\font minus
  \fontdimen4\font\relax}
\providecommand{\BIBforeignlanguage}[2]{{%
\expandafter\ifx\csname l@#1\endcsname\relax
\typeout{** WARNING: IEEEtran.bst: No hyphenation pattern has been}%
\typeout{** loaded for the language `#1'. Using the pattern for}%
\typeout{** the default language instead.}%
\else
\language=\csname l@#1\endcsname
\fi
#2}}
\providecommand{\BIBdecl}{\relax}
\BIBdecl

\bibitem{ref_dt}
H.~Xu \emph{et~al.}, ``Smart mobility in the cloud:
  Enabling real-time situational awareness and cyber-physical control through a
  digital twin for traffic,'' \emph{IEEE Trans. Intell. Transp. Syst.},
  vol.~24, no.~3, pp. 3145--3156, 2023.

\bibitem{ref_sumo}
P.~A. Lopez \emph{et~al.}, ``Microscopic traffic simulation using {SUMO},'' in \emph{Proc. 21st IEEE
  Int. Conf. Intelligent Transportation Systems (ITSC)}, 2018, pp. 2575--2582.

\bibitem{ref_krauss}
S.~Krau{\ss}, ``Microscopic modeling of traffic flow: Investigation of
  collision free vehicle dynamics,'' Ph.D. dissertation, Universit\"at zu
  K\"oln, 1998.

\bibitem{ref_lc2013}
J.~Erdmann, ``Lane-changing model in {SUMO},'' in \emph{Proc. SUMO 2014:
  Modeling Mobility with Open Data}, vol.~24.\hskip 1em plus 0.5em minus
  0.4em\relax Berlin, Germany: DLR, 2014, pp. 77--88.

\bibitem{ref_huainan2022}
Y.-P. Fl{\"o}tter{\"o}d and M.~Behrisch, ``Online calibration with {SUMO} for
  network-wide traffic and emission monitoring: Case study {ITS} Huainan,'' in
  \emph{Proc. SUMO User Conference}, ser. SUMO Conference Proceedings,
  vol.~2.\hskip 1em plus 0.5em minus 0.4em\relax TIB Open Publishing, 2022, pp.
  1--11.

\bibitem{ref_ga}
D.~E. Goldberg, \emph{Genetic Algorithms in Search, Optimization, and Machine
  Learning}.\hskip 1em plus 0.5em minus 0.4em\relax Reading, MA, USA: Addison-Wesley, 1989.

\bibitem{ref_spsa}
J.~C. Spall, ``Multivariate stochastic approximation using a simultaneous
  perturbation gradient approximation,'' \emph{IEEE Trans. Autom. Control},
  vol.~37, no.~3, pp. 332--341, 1992.

\bibitem{ref_cmaes}
N.~Hansen and A.~Ostermeier, ``Completely derandomized self-adaptation in
  evolution strategies,'' \emph{Evolutionary Computation}, vol.~9, no.~2, pp.
  159--195, 2001.

\bibitem{ref_surrogate2}
D.~R. Jones, M.~Schonlau, and W.~J. Welch, ``Efficient global optimization of
  expensive black-box functions,'' \emph{Journal of Global Optimization},
  vol.~13, no.~4, pp. 455--492, 1998.

\bibitem{ref_surrogate1}
B.~Shahriari, K.~Swersky, Z.~Wang, R.~P. Adams, and N.~de~Freitas, ``Taking the
  human out of the loop: A review of {Bayesian} optimization,'' \emph{Proc.
  IEEE}, vol. 104, no.~1, pp. 148--175, 2016.

\bibitem{ref_tpe}
J.~Bergstra, R.~Bardenet, Y.~Bengio, and B.~K\'egl, ``Algorithms for
  hyper-parameter optimization,'' in \emph{Advances in Neural Information
  Processing Systems (NeurIPS)}, vol.~24, 2011, pp. 2546--2554.

\bibitem{ref_pracbo}
J.~Snoek, H.~Larochelle, and R.~P. Adams, ``Practical {Bayesian} optimization
  of machine learning algorithms,'' in \emph{Advances in Neural Information
  Processing Systems (NeurIPS)}, vol.~25, 2012, pp. 2951--2959.

\bibitem{ref_smac2011}
F.~Hutter, H.~H. Hoos, and K.~Leyton-Brown, ``Sequential model-based
  optimization for general algorithm configuration,'' in \emph{Proc. Learning
  and Intelligent Optimization (LION)}, ser. LNCS, vol. 6683, 2011, pp.
  507--523.

\bibitem{ref_bocal}
S.~Agriesti, V.~Kuzmanovski, J.~Hollm{\'e}n, C.~Roncoli, and B.~Nahmias-Biran,
  ``A {Bayesian} optimization approach for calibrating large-scale
  activity-based transport models,'' \emph{IEEE Open J. Intell. Transp. Syst.},
  vol.~4, pp. 740--754, 2023.

\bibitem{ref_aical}
P.~Manglano-Redondo, A.~Paricio-Garcia, and M.~A. Lopez-Carmona,
  ``Spatio-temporal {AI} modeling for urban traffic calibration: A {SUMO}-based
  approach,'' in \emph{Proc. SUMO User Conference}, ser. SUMO Conference
  Proceedings, vol.~6.\hskip 1em plus 0.5em minus 0.4em\relax TIB Open
  Publishing, 2025, pp. 33--50.

\bibitem{ref_bo4mob}
S.~Ryu, D.~Kwon, S.~Choi, A.~Deshwal, S.~Kang, and C.~Osorio, ``{BO4Mob}:
  {Bayesian} optimization benchmarks for high-dimensional urban mobility
  problem,'' in \emph{Advances in Neural Information Processing Systems
  (NeurIPS)}, vol.~38, 2025, pp. 24\,282--24\,316, Datasets and Benchmarks Track.

\bibitem{ref_calibration}
B.~Ciuffo and V.~Punzo, ``No Free Lunch theorems applied to the calibration
  of traffic simulation models,'' \emph{IEEE Trans. Intell. Transp. Syst.},
  vol.~15, no.~2, pp. 553--562, 2014.

\bibitem{ref_punzo2021}
V.~Punzo, Z.~Zheng, and M.~Montanino, ``About calibration of car-following
  dynamics of automated and human-driven vehicles: Methodology, guidelines and
  codes,'' \emph{Transportation Research Part C: Emerging Technologies}, vol.
  128, 2021, Art.~no.~103165.

\bibitem{ref_schrader2024}
M.~Schrader, A.~Karnik, A.~Hainen, and J.~Bittle, ``Calibrating car-following
  models using {SUMO}-in-the-loop and vehicle trajectories from roadside radar:
  Calibrating {CF} model parameters,'' in \emph{Proc. SUMO User Conference},
  ser. SUMO Conference Proceedings, vol.~5.\hskip 1em plus 0.5em minus
  0.4em\relax TIB Open Publishing, 2024, pp. 209--233.

\bibitem{ref_abuamer2024}
I.~M.~A. Abuamer and C.~M.~J. Tamp{\`e}re, ``Traffic microsimulation for smart
  cities: Investigating the impact of objective function formulation on
  calibration efficiency,'' \emph{IET Smart Cities}, vol.~6, no.~4, pp.
  276--290, 2024.

\bibitem{ref_zhang2017}
C.~Zhang, C.~Osorio, and G.~Fl{\"o}tter{\"o}d, ``Efficient calibration
  techniques for large-scale traffic simulators,'' \emph{Transportation
  Research Part B: Methodological}, vol.~97, pp. 214--239, 2017.

\bibitem{ref_sha2020}
D.~Sha, K.~Ozbay, and Y.~Ding, ``Applying {Bayesian} optimization for
  calibration of transportation simulation models,'' \emph{Transportation
  Research Record}, vol. 2674, no.~10, pp. 215--228, 2020.

\bibitem{ref_moriconi2020}
R.~Moriconi, M.~P. Deisenroth, and K.~S. Sesh~Kumar, ``High-dimensional
  {Bayesian} optimization using low-dimensional feature spaces,'' \emph{Machine
  Learning}, vol. 109, no.~9--10, pp. 1925--1943, 2020.

\bibitem{ref_lancer2023}
A.~Zharmagambetov, B.~Amos, A.~Ferber, T.~Huang, B.~Dilkina, and Y.~Tian,
  ``Landscape surrogate: Learning decision losses for mathematical optimization
  under partial information,'' in \emph{Advances in Neural Information
  Processing Systems (NeurIPS)}, vol.~36, 2023, pp. 27\,332--27\,350.

\bibitem{ref_fhwa_moe}
\BIBentryALTinterwordspacing
R.~Dowling, ``Definition, interpretation, and calculation
  of traffic analysis tools measures of effectiveness,'' U.S. Department of
  Transportation, FHWA, Tech. Rep. FHWA-HOP-08-054, 2007, Traffic Analysis Toolbox
  Vol.~VI. [Online]. Available:
  \url{https://ops.fhwa.dot.gov/publications/fhwahop08054/}
\BIBentrySTDinterwordspacing

\bibitem{ref_hcm}
{Transportation Research Board}, \emph{Highway Capacity Manual}, 6th~ed.\hskip
  1em plus 0.5em minus 0.4em\relax Washington, DC, USA: Transportation Research
  Board, 2016.

\bibitem{ref_lhs}
M.~D. McKay, R.~J. Beckman, and W.~J. Conover, ``A comparison of three methods
  for selecting values of input variables in the analysis of output from a
  computer code,'' \emph{Technometrics}, vol.~21, no.~2, pp. 239--245, 1979.

\bibitem{ref_turbo2019}
D.~Eriksson, M.~Pearce, J.~Gardner, R.~D. Turner, and M.~Poloczek, ``Scalable
  global optimization via local {Bayesian} optimization,'' in \emph{Advances in
  Neural Information Processing Systems (NeurIPS)}, vol.~32, 2019, pp.
  5496--5507.

\bibitem{ref_ensemble}
B.~Lakshminarayanan, A.~Pritzel, and C.~Blundell, ``Simple and scalable
  predictive uncertainty estimation using deep ensembles,'' in \emph{Advances
  in Neural Information Processing Systems (NeurIPS)}, vol.~30, 2017, pp. 6402--6413.

\bibitem{ref_sind}
Y.~Xu \emph{et~al.}, ``{SIND}: A drone dataset at signalized intersection in China,'' in
  \emph{Proc. 25th IEEE Int. Conf. Intelligent Transportation Systems (ITSC)},
  2022, pp. 2471--2478.

\bibitem{ref_ute}
R.~Feng, H. Zhu, N.~N. Sze, S.~Wang, and Z.~Li, ``Ubiquitous traffic eyes:
  Trajectory dataset focus on multiple traffic states and state transition on
  urban expressways,'' \emph{Transportation Letters}, vol.~18, no.~2, pp.
  446--462, 2026.

\bibitem{ref_wilcoxon}
F.~Wilcoxon, ``Individual comparisons by ranking methods,'' \emph{Biometrics
  Bulletin}, vol.~1, no.~6, pp. 80--83, 1945.

\end{thebibliography}
\else

\fi

\end{document}